\documentclass[seceq]{ptptex}

\notypesetlogo                       

\title{Transfer of Life-Bearing Meteorites from Earth to Other Planets
}

\author{
Tetsuya {\textsc Hara}\footnote{E-mail:hara@cc.kyoto-su.ac.jp}, 
Kazuma {\textsc Takagi}, 
and Daigo {\textsc Kajiura}Firstname 
}

\inst{
Department of Physics, Kyoto Sangyo University, Kyoto 603-8555, Japan}

\abst{
The probability is investigated that the meteorites originating on Earth are transferred to other planets in our Solar System 
and to extra solar planets. 
 We take the collisional Chicxulub crater event, and material that was ejected as an example of Earth-origin meteors.  
 If we assume the appropriate size of the meteorites as 1cm in diameter, the number of meteorites to reach 
 the exoplanet system (further than 20 ly) would be much greater than one.  We have followed the ejection 
 and capture rates estimated by Melosh (2003) and the discussion by Wallis and Wickramasinghe (2004).
If we consider the possibility that the fragmented ejecta (smaller than 1cm) are accreted to comets and other icy bodies, 
then buried fertile material could make the interstellar journey throughout Galaxy.  If life forms inside remain viable, 
this would be evidence of life from Earth seeding other planets.
  We also estimate the transfer velocity of the micro-organisms in the interstellar space.  In some assumptions, it could be estimated 
  that, if life has originated $10^{10}$\ years ago anywhere in our Galaxy as theorized by Joseph and Schild (2010a, b), 
  it will have since propagated throughout our Galaxy and could have arrived on Earth by 4.6 billion years ago.  Organisms disperse. 
}

\begin{document}

\maketitle

\section{Introduction}


A number of scientists now believe that micro-organisms can be transferred to and from various planets and moons, 
including from Earth 
to other worlds (Joseph and Schild 2010a,b ; Napier and Wickramasinghe 2010; Wainwright et al., 2010).  
To determine the probability of the transfer of viable organisms between planets, we have used Earth as the origin of these life-bearing 
rocks and have determined the likelihood that viable life could be deposited on those stellar objects in this Solar System 
 which are believed capable of sustain life, i.e. Enceladus, Europa, Ceres and dwarf planets Eris.  Recently it has been reported 
 that the detection of the super-Earth planets in the Gl 581 system which reside at the warming edge of 
  the habitable zone of the star (Udry et. al. 2007).
Therefore, we also investigate the probability of successful transfer to extra solar planets, such as Gl 581. 
  The propagation distances of life-bearing rocks are also estimated, and we have determined that under some circumstances,
  life originating in one stellar system could propagate throughout Galaxy.

\section{Seeding Other Planets With Life}

Super-Earth planets have been recently detected, including in the Gl 581 system where planet orbits are at the warming edge of 
the habitable zone of the star (Udry et al. 2007).  Whether this planet or other super-Earth's may harbor life or not is 
completely unknown.  
On the other hand, if microbial life were to be deposited on a super-Earth through mechanisms of panspermia 
(Joseph and Schild 2010a,b, Napier and Wickramasinghe 2010),
 then it could be predicted that these microbes flourish and reproduce.

 The only planet, which we know has life, is Earth.  Therefore, Earth would be a likely source to seed other planets with life.  
 This could take place 
 following solar storms which eject microbes from outer atmosphere into space, or from bolide impacts which eject stones, 
 rocks, and oceans of water into space (Joseph and Schild 2010a, b;
  Napier and  Wickramasinghe 2010).  Naturally, those meteors, asteroids or comets which strike with the strong force, would eject 
  the most material into space.  Thus it could be prediced that the asteroid or meteor
    which struck this planet 65 million years ago, and which created the Chicxulub crater (Alvarez. et al. 1979) would have ejected 
    substantial amount of rock, soil, 
    and water into space, some of which would have fallen onto other planets and moons, including stellar bodies outside our 
    stellar system, Kuiper belt objects, Oort cloud objects, and possibly 
    extrasolar planets.  That meteorite was estimated to be about 10km in diameter (Alvarez, et al. 1979, Bralower et al. 1998).     　　

 Here, we investigate the transfer probability of Earth origin rocks to our Solar System.  We put parameters as following  
 that $N_0$ rocks are ejected from Earth,
    $ s $(cm) is the distance to the object,  and the cross section of the rock capture by the object is $\sigma$ (cm$^2$).  
Then the impact rock number is estimated that $N_0$ times $\sigma$ over the surface of sphere of radius $s$ as 
%
\[
 N_{impact}=N_0 \left(\frac{\sigma}{4 \pi s^2}\right)
 \]
%

  When the Chicxulub meteorite collided to the earth, it could be estimated that almost the same amount mass 
could be ejected from Earth, where we have taken the optimistic value (Wallis and Wickramasinghe, 2004).  
Then it is assumed that the ejected mass 
from Earth is $f_1 \times  M_0$, where $ M_0$ is the mass of the Chicxulub meteorite and
the factor $f_1 (\sim 0.3)$ denotes the fraction of the mass ejected from Earth.  Taking that the mean diameter of rocks is $r_1$ (cm) and 
the estimated diameter of the Chicxulub meteorite is $R_1$, the number $N_0$ of ejected 
rocks from Earth is estimated as $f_1$ times  of the cubic of $(R_1/r_1)$.

If we take $R_1$ and $r_1$ as 10(km) and 1(cm), the number $N_0$ of ejected rocks is the order of 
\[
N_0=f_1 \left(\frac{R_1}{r_1}\right)^3 \sim 3\times 10^{17}
\left( \frac{f_1}{0.3}\right) \left( \frac{R_1}{10  \hspace{0.1cm} \rm{km}} \right)^3 
\left(\frac{r_1}{1 \hspace{0.1cm}  \rm{cm}}\right)^{-3} .
\]
This is a rather crude approximation.  To be precise, we have to include the size distributions of rock fragments.  
Here we only tentatively want to estimate the number of rocks.  If we take $r_1  \sim 1$mm, the above value has increased factor $\sim   10^3$.
  However the cosmic radiation in space will damage the micro-organism within the fragments of size  $\sim 1$ mm, 
  unless it is covered by molecules such as ice or other elements.

The distance to the interesting objects within our Solar System is the order of astronomical unit.  
So we take the representative value as $s \sim$  1AU $\sim   1.5\times 10^8$ {\rm km}.  
The problem is the cross section $\sigma$.  So we consider the following two models.
\vspace{0.3cm}

Model A :  The cross section $\sigma$  for the direct collision to the object is 
the order  $\sigma \sim  \pi R_0^2 $ where $R_0$ is the radius of the object.  
Then the number of impact rocks is estimated for the case $R_0 \sim   10^ 3$ {\rm km} and $s \sim   1$ AU
as 
\[
 N_A \sim   N_0 \left(\frac {\sigma}{4\pi s^ 2}\right)
 \]
 \[ \sim  3\times 10^6 \left(\frac{f_1}{0.3}\right)\left(\frac{R_1}{10{\rm km}}\right)^3
 \left(\frac{r_1}{1\rm{cm}}\right)^{-3}\left(\frac{R_0}{10^3 {\rm km}}\right)^2
 \left(\frac{s}{1{\rm AU}}\right)^{-2}.
 \]
This model corresponds to the high velocity case of ejected rocks.
\vspace{0.3cm}

Model B :  After rocks are ejected from Earth, they are orbiting around Earth and then ejected to orbits around Sun 
through swing-by.  If rocks could be decelerated by gravitational interaction, rocks are captured to objects.  
After a few My, some fraction of rocks could fall into objects.  The gravitational infall to the object 
could be inferred by the gravitational accretion radius $R_g \sim G m_0/v_0^2$ where $m_0$ is the mass of the object. 
Then the cross section $\sigma$ is estimated as $ \sigma \sim   R_g ^2 $ which is proportional to the mass square $m_0^2$.  
As the dominant planet is Jupiter, the infalling rate is roughly proportional to $(m_0/M_J)^ 2$  where $M_J$ is Jupiter's mass.  
Then the number of falling rocks is estimated for the case of $m_0 \sim 10^{20}$ kg as 
\[
N_B \sim  N_0\left(\frac{m_0}{M_J}\right)^2 
\]
\[\sim 10^3\left(\frac{f_1}{0.3}\right) \left(\frac{R_1}{10{\rm km}}\right)^3
\left(\frac{r_1}{1{\rm cm} }\right)^{-3}\left(\frac{m_0}{10^{20} {\rm kg}}\right)^2
 \left(\frac{M_J}{2\times 10^{27}  {\rm kg}}\right)^{-2}.
\]
This model corresponds to the low velocity case of ejected rocks. The values for Enceladus, Europa, Ceres, Eris, Moon and Mars are 
presented in the Table.

\begin{table}
\vspace{0.4cm}
\begin{center}
 \begin{tabular}{|l|r|r|r|r|r|} \hline \hline
  \rule[0.2cm]{0.0cm}{0.3cm}   \hspace{0.5cm} Object \hspace{0.3cm}  &   \hspace{0.2cm} s (AU) \hspace{0.2cm}
 & \hspace{0.3cm}$R_0$({\rm km})\hspace{0.3cm} &\hspace{0.1cm} $m_0$ (kg) \hspace{0.1cm} & 
 \hspace{0.6cm} $N_A$  \hspace{0.1cm} & \hspace{0.6cm} $N_B$  \hspace{0.1cm}  \\  [4pt] \hline \hline
 \rule[0.2 cm]{0.0 cm}{0.2cm}  
   \hspace{0.4cm}  Enceladus \hspace{0.3cm} & 10 \hspace{0.2cm} & $2.5\times 10^2$ \hspace{0.1cm} & $ 7\times 10^{19}$\hspace{0.1cm} 
   & $ 2\times 10^{3}$ \hspace{0.1cm} & $ 5\times 10^{2}$\hspace{0.1cm}\\  
   \hspace{0.5cm}  Europa   &  5 \hspace{0.2cm}   &   $ 1.6\times 10^{3} $  \hspace{0.1cm} &   $5\times 10^{22}\hspace{0.1cm}$  
   & $ 3\times 10^{5}$ \hspace{0.1cm} & $ 3\times 10^{8}$\hspace{0.1cm} \\
   \hspace{0.5cm}  Ceres   &   3  \hspace{0.2cm}   &  $ 5\times 10^{2} $ \hspace{0.1cm} &   $9\times 10^{20} \hspace{0.1cm} $ 
   & $ 10^{5}$ \hspace{0.1cm} & $10^{5} \hspace{0.1cm}$\\  
   \hspace{0.5cm}  Eris     &   100  \hspace{0.2cm}  &  $ 1.2\times 10^{3}$  \hspace{0.1cm}  &  $ 2\times 10^{22} \hspace{0.1cm}$ 
   & $ 4\times 10^{2} \hspace{0.1cm} $  & $4\times 10^{7}$\hspace{0.1cm} \\
   \hspace{0.5cm}  Moon   &  $ \hspace{0.2cm}3\times 10^{-3} $  &  $1.7\times 10^{3} \hspace{0.1cm} $  & $7\times 10^{22}\hspace{0.1cm}$  
   & $ 10^{12} \hspace{0.1cm}$ & $5\times 10^{8}\hspace{0.1cm}$ \\ 
   \hspace{0.5cm}  Mars   &  $ 1.5 \hspace{0.3cm}$  &  $3.4\times 10^{3} \hspace{0.1cm} $ & $6\times 10^{23}\hspace{0.1cm}$  
   & $ 2\times 10^{7}\hspace{0.1cm}$ & $4\times 10^{10}\hspace{0.1cm}$ \\ \hline 
\end{tabular}
\end{center}
\vspace{0.3cm}
\caption{The values of $s, R_0, m_0, N_A, N_B$ for Enceladus, Europa, Ceres, Eris, Moon and Mars} 
\end{table}
\vspace{0.6cm}

For every object, the number $N_A$ and $N_B$ are much greater than one. 
 Although it is uncertain how rocks enter the presumed sea under the surface, for example, of Enceladus and Europa, 
 the probability may be high that micro-organisms transferred from Earth would be adapted and growing there.  
 The orbital calculations of meteorite transfer among planets within our Solar System are estimated by Melosh (2003).

\section{Probability of reaching Gl 581 and extra solar planets}

To extend the above consideration to the extra solar planets is almost straightforward.  
We introduce the factor $f_2 (\sim  0.3)$ which denote the fraction of rocks ejected from our Solar System. 
As the distance to Gl 581 is 20 light years, we take the representative value for $s$ as $s \sim  10^{19} ({\rm cm} )$. 
The problem is the cross section $\sigma$. So we consider the following model A and C, where model A is the almost 
the same in section 2.
\vspace{0.3cm}

Model A :  The cross section  $\sigma$ for the direct collision to the planet in the Gl 581 system is of 
the order $\sigma \sim  \pi r^2  \sim  \pi(m /m_{\oplus})^{2/3}r_{\oplus}^2 $ where $m$ and $m_{\oplus}$ are the mass of 
the planet 
and Earth, respectively. Then the cross section factor becomes 
$\sigma/(4\pi s^2) \sim  (m /m_{\oplus})^{2/3}(r_{\oplus}/s)^2/4 \sim  3 \times 10^{-21}(m /5m_{\oplus})^{2/3}(s /10^{19}{\rm cm} )^{-2}, 
$
and the impact number becomes as 
\[
 N_{impact} \sim N_0\left(\frac{\sigma}{4 \pi s^2}\right) 
\]
\[ \sim 3 \times 10^{-4}\left(\frac{f_1 f_2 }{0.1}\right)\left(\frac{R_1}{10{\rm km}}\right)^3
 \left(\frac{r_1}{1{\rm cm} }\right)^{-3}\left(\frac{m }{5 m_{\oplus}}\right)^{2/3}
 \left(\frac{s }{10^{19} {\rm cm} }\right)^{-2} .
\]
The probability for the direct collision is small.
\vspace{0.3cm}

Model C:  The cross section $\sigma$ could be enlarged including the effect of the gravitational interaction 
such as swing-by.  
If rocks could be deaccelerated by gravitational interaction, rocks are captured to the stellar system.  
Although the velocity dependence of the cross section is pointed out by Melosh (2003), it is difficult to include 
this effect here.  
Then we simply assume the cross section as $\sigma \sim f_3 \pi  R^2$, where $R$ is the orbit radius of the planet and 
the uncertainty factor $f_3$ is included.  The number of impacted rocks, $N_{impact}$, becomes 
\[
N_{impact} \sim  N_0 \left(\frac{f_2 f_3}{4 }\right) \left(\frac{R}{s}\right)^2 
\sim 10^4\left(\frac{N_0f_2}{10^{17}}\right)\left(\frac{f_3}{0.1}\right)
\left( \frac{R}{1{\rm AU}} \right)^2\left(\frac{s}{10^{19}{\rm cm}}\right)^{-2}.  
\eqno(1)\]
Due to the estimation of numerical simulations by Melosh (2003), the factor $f_1$, and $f_2$ are roughly 0.3. 
We take $f_3$ tentatively 
as $\sim 0.1$.  Even though there are many uncertainty factors, the probability could increase by considering 
the small rocks which are 
smaller than 1{\rm cm} .
If we consider the possibility that the fragmented ejecta (smaller than 1{\rm cm} ) are accreted to comets and other icy bodies 
in the 
'Edgeworth-Kuiper Belt', the securely buried fertile material could make the interstellar journey through Galaxy 
(Wallis and Wickramasinghe, 2004). 
The above estimated number $N_{impact}$ is the captured number of rocks in the Gl 581 system. If we consider 
the falling probability $f_4$ 
to the appropriate planet and the landing probability $f_5$ to the appropriate circumstances of the planet, the probability for 
the proliferation of the life must be decreased. Then the numbers of rocks for the proliferation becomes 
\[
N_{proli} \sim 10^2 \left(\frac{f_1f_2f_3}{10^{-2}}\right)\left(\frac{f_4f_5}{10^{-2}}\right)\left(\frac{R_1}{10{\rm km}}\right)^3
\left(\frac{r_1}{1{\rm cm} }\right)^{-3}
\]
\[\times \left(\frac{R }{1{\rm AU}}\right)^2 
\left(\frac{s}{10^{19}{\rm cm} }\right)^{-2}.  
\]
If we take the mean velocity of meteorites in the interstellar space as $\sim$ 10 {\rm km}/s, the elapsed time to travel to GI 581 system is
\[ 
T \sim  \left(\frac{s}{v }\right) \sim  10^6 \left(\frac{s}{20 {\rm ly}}\right)\left(\frac{10 {\rm km s^{-1}}}{v}\right) \  {\rm years}.
\]
Then the time to be ejected from our Solar System through swing-by (several Million years) and the orbiting time 
to fall in the planet 
through swing-by after captured to Gl 581 system (several Million years) are longer than the travel time to Gl 581 system.

\section{Transfer Distance and Velocity of Rocks with Micro-Organisms}

In this section, we estimate the distance and velocity of rocks with micro-organism through the interstellar space.
\vspace{0.3cm}

Model I:  The number of ejected rocks from our Solar System is 
\[
N_{ej} \sim  10^{17} \left(\frac{f_1f_2}{0.1}\right)\left(\frac{R_1}{10{\rm km}}\right)^3 \left(\frac{r_1}{1{\rm cm} }\right)^{-3}.  
\]

To apply the estimate to general stellar systems, we change the size of the system from $\sim  1$ AU to $\sim  10$ AU 
where the orbits of Jupiter and/or Saturn like planets are considered.  If the radius of the planet orbit is taken as $R$ , 
the cross section $\sigma$  is given by  $\sigma \sim \pi R^2$.  The number of accumulated rocks to the system is 
\[
N_{impact} \sim N_{ej}\left(\frac{\pi R^2}{4\pi L^2}\right), 
\]
where $L$ is the distance between the origin and the system.  For the propagation of the micro-organism, $N_{impact}$ 
must be greater 
than $N_{crit} (\sim 1)$.  From the criterion $(N_{impact} \geq N_{crit})$, the distance $L$ is limited as
\[
L \leq L_{crit} \sim \left(\frac{N_{ej}}{N_{crit}}\right)^{1/2}\left(\frac{R}{2 }\right) \sim 
3 \times 10^4 \left(\frac{N_{ej}}{10^{17}}\right)^{1/2}(N_{crit})^{-1/2} \left(\frac{R}{10{\rm AU}}\right) \ {\rm ly}.
\]
The propagation time $T$ to the $L_{crit}$ is given by
\[
T \sim \left(\frac{L_{crit}}{v_{mean}}\right) \sim 10^9 \left(\frac{L_{crit}}{3 \times 10^4 {\rm ly}}\right) 
\left(\frac{10 {\rm km \ s^{-1}}}{v_{mean}}\right)\  {\rm years},
\]
where we take the mean velocity of rocks as 10 ${\rm km s^{-1}}$.  Then, by $10^9$ years, rocks could reach each stellar system 
within the distance $3\times 10^4$ light years. 
The above estimate is rather optimistic.  We must consider the uncertainty factor such as $f_3 , f_4$ and $f_5$.  
If we include these factors, the value of $N_{crit}$ must be greater than $10^3$.  Then the above critical length decreased to 
\[
L_{crit} \sim \left(\frac{N_{ej}}{N_{crit}}\right)^{1/2} \left(\frac{R}{2}\right) \sim 10^3 \left(\frac{N_{ej}}{10^{17}}\right)^{1/2} 
\left(\frac{10^3}{N_{crit}}\right)^{1/2} \ {\rm ly}.
\]
The propagation time $T$ to $L_{crit}$ is given by
\[
T \sim  \left(\frac{L_{crit}}{v }\right) \sim 3\times 10^7\left(\frac{L_{crit}}{10^3 \ {\rm ly}}\right)\left(\frac{10 {\rm km \  s^{-1}}}{v}\right)\   {\rm years}.  
\]
Then, by $3\times 10^7$ years, the many rocks $(\sim 10^3)$ could reach each stellar system within the distance 
$10^3$ light years.  
To estimate the longer distance than $L_{crit}$, we consider the following model.
\vspace{0.3cm}

Model II:  The Chicxulub crater event is 65 My ago and such event is happened roughly per every 100 My, 
which is consistent of the accretion rate of crater forming bodies ($\sim 10^{11}$g/y, Sephton (2003)).  
Then the mean number of ejected rocks per year $N_{mean}$ is estimated as $N_{mean} \sim  N_{ej}/(10^8{\rm y}) \sim 10^9/{\rm y}.$
 The number $N_{acc} (t)$ of accumulated rocks in the system of the distance $L$ after $t$ year is
\[
N_{acc}(t) \sim N_{mean} \times t \times \pi R^2/(4\pi L^2).
\]
Then the distance $(L_{crit}(t))$ which satisfies the criterion that $N_{acc}(t)$ is greater 
than $N_{crit}( \sim 10^3)$ is given  by 
\[
L_{crit}(t) \sim  (N_{mean} \times t \times \left(\frac{R^2}{4 N_{crit}}\right))^{1/2}  
\]
\[
\sim 10^3 \left(\frac{N_{mean}}{10^9}\right)^{1/2} \left(\frac{t}{10^8{\rm y}}\right)^{1/2}\left(\frac{10^3}{N_{crit}}\right)^{1/2}
\left(\frac{R}{10{\rm AU}}\right)\  {\rm ly}.
\]
As the radius of Galaxy $(R_G)$ is about $5\times 10^4$ ly, it takes $3\times 10^{11}$ years for $L_{crit}(t) \geq R_G$, 
which is much greater than the age of our Galaxy. 
 \vspace{0.3cm}

 Model III:  If we assume that the propagated system becomes the place where the micro-organisms adapt, multiply and proliferate, 
 the system becomes the source of the rocks with micro-organisms.  If it takes $t_0$ time to propagate to 
 such a system with distance $L$ 
 and $t_1$ time to proliferate there enough, the propagation velocity $v_{prop}$ is
\[
 v_{prop} \sim  \left(\frac{L }{t_0 + t_1}\right) \sim  \left(\frac{10^3{\rm ly}}{10^8 \ {\rm y}}\right),
\]
where we take $ L \sim  L_{crit} \sim 10^3$ ly, $t_0 \sim 3 \times 10^7 $ years and  $t_1 \sim 10^8$ years, respectively.
\vspace{0.1cm}

１）If the multiplication factor $ m $ of such descendants for each generation is high $(m \gg 1)$ , the propagated distance 
$L_{prop}$ would be proportional to the time as
\[
 L_{prop}  \sim v_{prop} \times t . 
\]
It takes $5\times 10^9$ years for $L_{prop} \geq R_G$, which means that if origin of life has begun within our 
Galaxy $10^{10}$ years ago, 
it has propagated through Galaxy, as our Galaxy age is almost $1.3\times 10^{10}$ years.  The problem may be 
that many types of life 
which evolved differently from the same origin are falling to Earth nowadays. 
\vspace{0.1cm}

2) If the multiplication factor $m$ of such descendants for each generation is not high $(m\simeq 1)$ and they eject rocks 
with micro-organism only one time after $t_1$, the propagated distance $L_{prop}$ would be proportional to 
the square root of time 
for its random walk property, as
\[
 L_{prop}\sim  L_{crit}\times \left(\frac{t}{10^8 y}\right)^{1/2}.
\]
If we take $L_{crit} \sim 10^3 {\rm ly}$, it takes $2.5\times 10^{12}$ years for $L_{prop} \geq R_G$ .  
It means that, if origin of life has begun within our Galaxy $10^{10}$ years ago, 
it has propagated only $10L_{crit}\sim 10^4 {\rm ly}$.  
If there are $X \sim 25$ sites where the life began $10^{10}$ years ago in our Galaxy, the propagated surface is about 
the same of our Galaxy by the equation 
\[
 \pi (10L_{crit})^2 X \sim \pi R_G^2.
\]
Then the probability is almost one that our solar system is visited by the micro-organisms originated in extra solar system.
 
\section{Conclusions and Discussion}

Although there are many uncertain factors, the probability of rocks originated from Earth to reach nearby 
star system is not so small.  
The rough estimation is given by Eq. (1) as
\[N_{impact} \sim 10^4\left(\frac{N_0f_2}{10^{17}}\right)\left(\frac{f_3}{0.1}\right)
\left(\frac{R}{1{\rm AU}}\right)^2
\left( \frac{10^{19} \rm{cm}}{s} \right)^2
\eqno(1)
\]

Although it is not certain that the micro-organisms within the size 
$(\leq 1{\rm cm})$ of meteorites are still viable for several My under cosmic radiation in space, there is the possibility 
that the fragmented ejecta are covered by accreted molecules such as ice or other elements.  
It is pointed out the possibility that fragments are accreted by comets and other ice objects 
where the buried fertile material could endure the cosmic radiation (Wallis and Wickramasinghe, 2004).  
Under these circumstances fragments could continue the interstellar journey and 
Earth origin meteorites could be transferred to Gl 581 system.  If we take it is viable, 
we should consider the panspermia theories more seriously. 
We estimate the transfer velocity of the 
micro-organisms among the stellar systems.  Under some assumptions, it could be estimated that if origin of life has begun 
$10^{10}$ years ago 
in one stellar system as estimated 
by Joseph and Schild (2010a, b), it could propagate throughout our Galaxy by $10^{10}$ years, 
and could certainly have reached Earth 
by 4.6 billion years ago (Joseph 2009), thereby explaining the origin of life on Earth.
\vspace{0.5cm}
\\
\\
\hspace{5cm} {\bf References}
\vspace{0.6cm}
\\
Alvarez, W., Alvarez, L.W., Asaro, F., Michel, H.V.(1979). Anomalous iridium levels at 
the Cretaceous/Tertiary boundary at Gubbio, Italy
: Negative results of tests for a supernova origin, in, 
In: Christensen, W.K.,  Birkelund, T. (Eds), Cretaceous/Tertiary Boundary Events Symposium, 
University of Copenhagen, v. 2, p. 69.
\vspace{0.3cm}
\\
Joseph R., Schild, R. (2010a). Biological Cosmology and the Origins of Life in 
the Universe Journal of Cosmology, 5, 1040-1090.
\vspace{0.3cm}
\\
Joseph R., Schild, R. (2010b). Origins, Evolution, and Distribution of Life in 
the Cosmos: Panspermia, Genetics, Microbes, and Viral Visitors From the 
Stars. Journal of Cosmology, 7, 1616-1670.
\vspace{0.3cm}
\\
Napier, W. M., Wickramasinghe, N. C. (2010). Mechanisms for Panspermia. 
Journal of Cosmology, 7, 1671-1691.
\vspace{0.3cm}
\\
Bralower, T. J., Paull, C. K., Leckie, R. M. (April 1998) 
The Cretaceous-Tertiary boundary cocktail: Chicxulub impact triggers margin collapse and extensive 
sediment gravity flows, Geology 26, 331-334.
\vspace{0.3cm}
\\
Melosh, H.J. (2003). Exchange of Meteorites (and Life?) Between Stellar Systems. Astrobiology, 3, 207-215. 
\vspace{0.3cm}
\\
Sephton, M.A. (2003). Origin of Life. In: Gilmour, I. Sephtons, M.A. (Eds), An Introduction to Astrobiology, 
Cambridge Univ. Press. Cambridge, pp 1-41.
\vspace{0.3cm}
\\
Udry, S. et. al. (2007). The HARPS search for southern extra-solar planets XI. Super-Earths (5 and 8 $M_{\oplus}$ ) 
in a 3-planet system , 
(astro-ph/0704.3841), Astronomy and Astrophys. 469, L43-L47.
\vspace{0.3cm}
\\
Wainwright, M., Alshammari, F., Alabri, K. (2010). Are Microbes Currently Arriving to Earth from Space? 
Journal of Cosmology, 7, 1692-1702.
\vspace{0.3cm}
\\
Wallis, M.K., Wickramasinghe, N.C. (2004). Interstellar transfer of planetary microbiota. MNRAS. 348, 52-61. 
%
\end{document}